\documentclass[prl,twocolumn,floatfix,showpacs]{revtex4}

\usepackage{amsmath,amsfonts} 
\usepackage{graphicx}
\usepackage{setspace}

\usepackage{dcolumn}   % Align table columns on decimal point
\usepackage{bm}        % bold math
\usepackage{subfigure}

\DeclareGraphicsExtensions{.eps}

\begin{document}

\title{Motion of a Viscoelastic 
Micellar Fluid Around a Cylinder: Flow and Fracture}

\author{Joseph R. Gladden\footnote[3]{Present address: Department of Physics and Astronomy, University of Mississippi, Oxford, MS.} 
and Andrew Belmonte}

\affiliation{The W.~G.~Pritchard Laboratories, Department of Mathematics,\\
Pennsylvania State University, University Park, PA 16802, USA}

\label{firstpage}

\date{May 18, 2006}

\begin{abstract}
We present an experimental study of the motion of a viscoelastic micellar material around a moving cylinder, which ranges in response from fluid-like flow to solid-like tearing and fracture, depending on the cylinder radius and velocity.
The observation of viscoelastic crack propagation driven by the cylinder indicates an extremely low tear strength, approximately equal to the steady state surface tension of the fluid.
At the highest speeds a driven crack is observed in front of the cylinder, propagating with a fluctuating speed equal on average to the cylinder speed, here as low as 5\% of the elastic wave speed in the medium.

\end{abstract}

\pacs{47.20.Gv, 47.57.-s, 46.50.+a}

\maketitle

%%%%%%%%%%%%%%%%%%%%%%%%%%%%%%%%%%%%%%%%%%%%%%%%%%
%%%%%%%%%%%%%%%%%%%%%%%%%%%%%%%%%%%%%%%%%%%%%%%%%%

By definition, the primary difference between a solid and a fluid is that the former resists an applied force by undergoing a finite deformation while the latter flows, or deforms, continually. Viscoelastic materials combine these aspects in their linear response, either as a viscous, lossy solid or as an elastic fluid \cite{larsonBK}.
A solid will fail under a large enough force, via different mechanisms such as crack formation and propagation (fracture) \cite{freundBK,lawn93,marder96,hauch99}. When viscoelastic materials fail, relaxation processes play a more prominent role, and the dynamics of the failure mechanisms are altered \cite{bonn98,gay99,hui03,persson05}. 
Studying fracture in soft materials holds the 
benefit of much lower sound speeds, which leads to slower and more accessible crack dynamics \cite{fineberg05, tanaka00}.  
Here we focus on the transition to solid-like failure in a concentrated viscoelastic micellar fluid;
how do different failure modes arise from the fluid-like dynamics at slower timescales and lower forces?

Viscoelastic micellar fluids are comprised of self-assembling surfactant aggregates, and are interesting as soft materials due to the substitution of hydrophobic `bonds' for covalent ones in the elastic material properties, a similarity they share with biological membranes \cite{israelBK}. 
At higher concentrations these fluids can be gel-like, although they still flow under small stress, due to the presence of micellar junctions instead of the chemical crosslinks of a polymer gel \cite{in99,koga05}. 
There are a variety of morphologies of these surfactant systems, including onion phases \cite{gulik96}, myelins \cite{buch}, and wormlike micellar fluids; the latter are known for novel rheological behavior \cite{larsonBK, 
lerouge00, liuandpine} and hydrodynamic instabilities \cite{ranjini,anand03,chen04,handzy04}, but to our knowledge there has been no study of crack propagation. In a recent experiment on sphere impact into micellar fluids, however, we did observe a transition from smooth to fractured cavity surface \cite{akers}.

\begin{figure}
\centering
\begin{center}
\includegraphics[width=3.2truein]{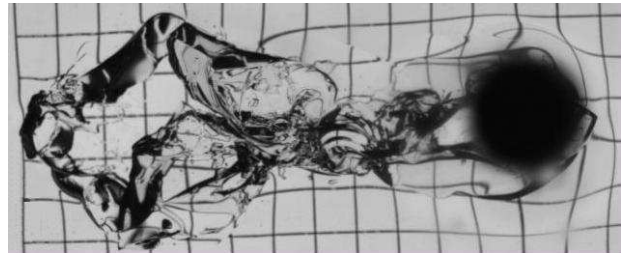} 
\end{center}
\vspace{-0.25cm}
\caption{ 
The initial result of quickly accelerating a cylinder from rest in a concentrated micellar fluid  ($d = 9.5$ mm, $U_0 = 77$ mm/s).}
\label{f-initialcut}
\end{figure}

In this Letter we present an experiment in which a cylinder is pulled through a shallow layer of concentrated micellar fluid, allowing us to pass from the flowing response of a viscous fluid to the tearing of a solid 
by controlling the timescale of the applied force.  At the limits of our experiment we observe an unsteady initial crack, followed by a ripped-up wake, and a leading crack ahead of the cylinder (see Figure \ref{f-initialcut}). In contrast to a cross-linked polymer gel, this material is ``self-healing", and the torn surface returns to a smooth, flat state 
after a few hours.

%%%%%%%%%%%%%%%%%

%%%%%%%%%%%%%%%%%

The concentrated micellar solution we study is well-known as a wormlike micellar fluid at lower concentrations: an aqueous solution of the surfactant cetyltrimethylammonium bromide (CTAB) and the organic salt sodium salicylate (NaSal), which facilitates the formation of long tubular ``wormlike'' aggregates \cite{larsonBK,israelBK,liuandpine}. 
We use 200 mM CTAB and 120 mM NaSal, concentrations for which we have found both fluid and solid-like behaviors \cite{IUTAMexico}. 
The CTAB and NaSal (obtained from Aldrich) are dissolved in filtered deionized water without further purification. Each solution is mixed separately at a necessarily high temperature ($\sim 70^\circ$C) \cite{handzy04}, then combined and mixed for several hours. The fluid thickens to a gel-like state as it cools, but is poured into the experimental channel or rheometer while still warm ($\sim 55^\circ$C); it sits for at least 24 hours before use.

%%%%%%%%%%%%%%%%%

\begin{figure}
\centering
\begin{center}
\includegraphics[width=2.6truein]{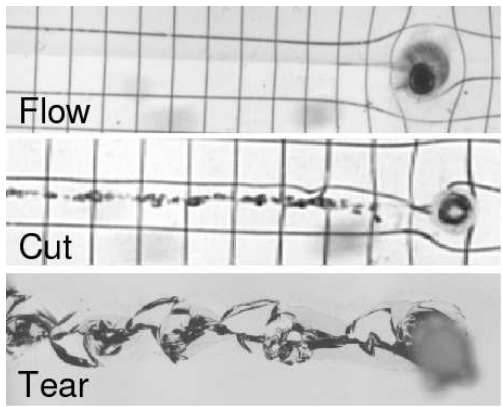}
\end{center}
\vspace{-0.25cm}
\caption{Three distinct responses of a 200/120 mM CTAB/NaSal micellar fluid to the moving cylinder ($d$=3.1, 1.8, and 7.9 mm respectively). The speeds are 2.8 mm/s (flow), 9.9 mm/s (cut), and 16 mm/s (tear). The view is along the cylinder axis, and the fluid depth is 2.0 cm.}
\label{fct_pics}
\end{figure} 

%%%%%%%%%%%%%%%%%

The rheology of this fluid was measured  at a fixed temperature of $22.8^{\circ}\,$C with an RFS-III strain rate controlled couette cell rheometer (Rheometric Scientific, now TA Instruments).  At low shear rates the flow is Newtonian, with a zero shear viscosity of $\eta_0 \simeq 1670$ Poise. At higher shear rates the fluid shear thins, and the stress reaches a roughly constant plateau, as is often seen in wormlike micellar fluids \cite{larsonBK,liuandpine,cates90}.
Dynamic rheology experiments were also performed and found to be well fitted by a Maxwell model for strain frequencies less than about 10 s$^{-1}$, yielding an elastic shear modulus of $G_0 \simeq$ 190 Pa \cite{IUTAMexico}. Both measurements indicated a relaxation time $\lambda \simeq  1.1$ s. The measured density was $\rho \simeq 1.1$ g/cc \cite{IUTAMexico}.

We have constructed a linear motion system by which a rigid rod (circular or square cross section) is pulled through the fluid at speed $U_0$.
Our setup was partly inspired by the mylar sheet cutting experiments done by Roman and coworkers \cite{roman03}.   
An acrylic box of dimensions 61 x 15 x 5 cm is filled with the warm (pourable) micellar fluid to a depth $H$ from 0.9 to 2.0 cm.  The box is mounted on a translation stage driven by a voltage controlled servo motor and ball screw with a range of 83 cm; the translation system maintains a fixed speed $U_0 \simeq$ 0-80 mm/s.  A cylinder of diameter $d$ is inserted into the fluid, $\sim$ 0.5 mm from the bottom.  The response of the material is documented with a Phantom v5.0 high speed digital video or a Nikon D70 digital SLR camera; the cylinder and the imaging system are at rest in the lab frame. The experiments were performed at room temperature (22.5 -- 24.5$^\circ$C). The flow is highly elastic (Deborah number $De  = \lambda U_0/d \simeq 0.5 - 11.5$) while remaining very viscous (Reynolds numbers $Re =\rho d U_0/\eta_0 \simeq 5 \times 10^{-4}$).

%%%%%%%%%%%%%%%%%%%%%%%%%%%%%%%%%%%%%%%%
\begin{figure}
\centering
\begin{center}
\includegraphics[height=2.15truein]{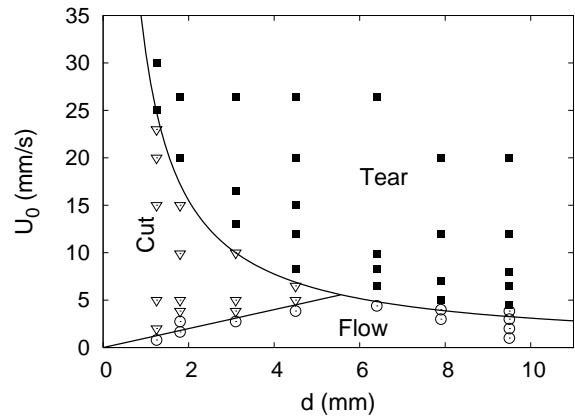}
\vspace{-0.3cm}
\end{center}
\caption{The phase diagram for a moving cylinder (speed $U_0$, diameter $d$) in a micellar fluid layer (depth 2.0 cm). The boundaries represent simple scaling relations (see text). 
}
\label{phase}
\end{figure} 
%%%%%%%%%%%%%%%%%%%%%%%%%%%%%%%%%%%%%%%%

We observe three distinct dynamic responses depending on the size $d$ and speed $U_0$ of the cylinder, as illustrated in Fig.~\ref{fct_pics}. In the \textit{Flow} state, the material smoothly moves around the cylinder and recombines behind it leaving a creased wake, but no trapped air bubbles. At higher speeds, a smooth cavity appears just behind the cylinder, a phenomenon also seen in viscous Newtonian fluids \cite{IUTAMexico}; the material still flows smoothly. In the \textit{Cut} state, the material still appears to flow around the cylinder, but the walls of the cavity are textured.  This results in air bubbles trapped as the walls recombine (Fig.~\ref{fct_pics}b). 
In contrast, the \textit{Tear} state is characterized by lateral splitting of the cavity walls, resulting in large tears with a characteristic fin shape (Fig.~\ref{fct_pics}).
In Fig.~\ref{phase} we show the material response for various cylinder velocities and diameters ($H = 2.0$ cm).   
For a large enough cylinder, the transition is directly from flow to tear.

\begin{figure}
\centering
{
\includegraphics[height=1.5truein]{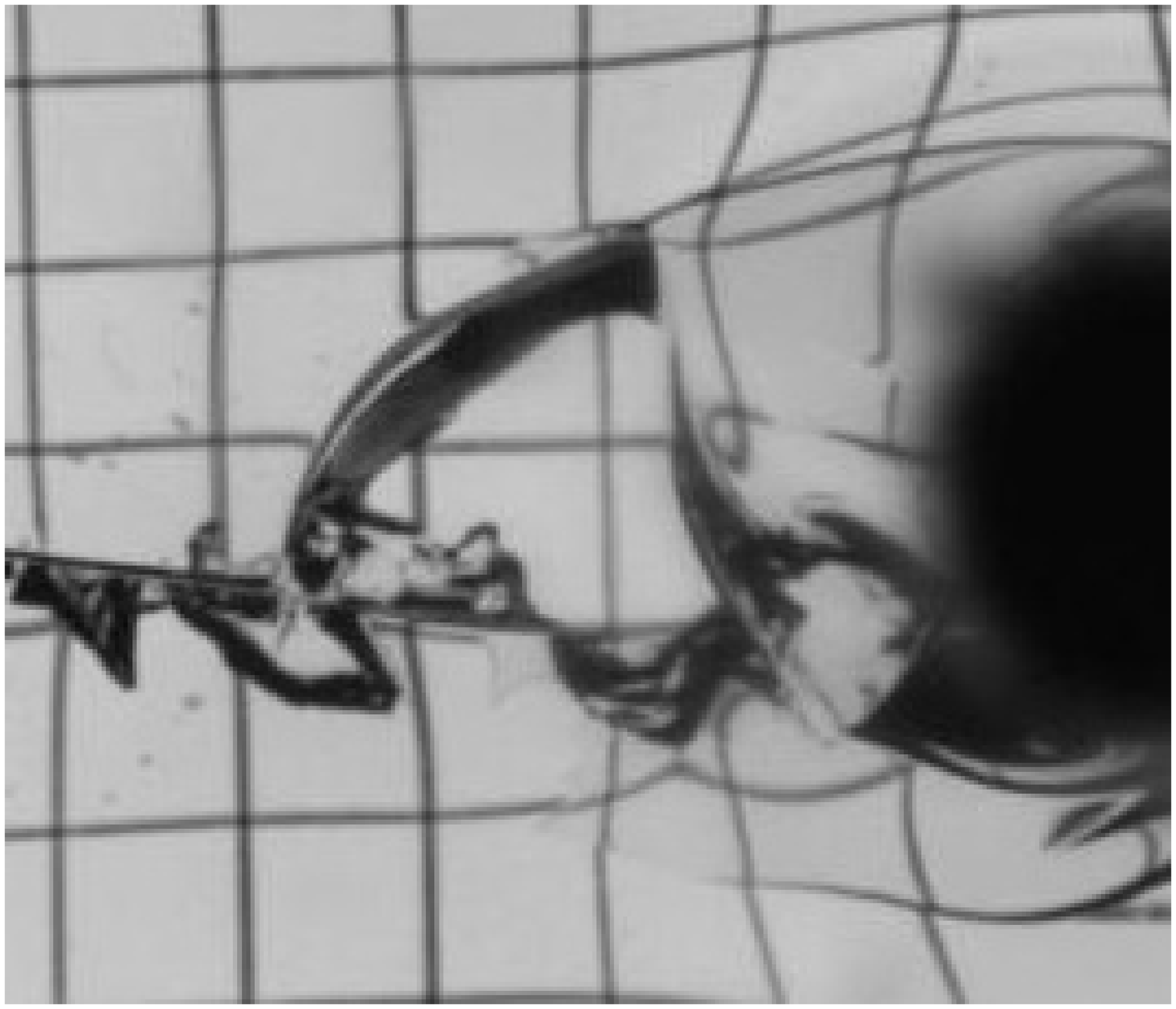}
\hspace{0.1cm}
\includegraphics[height=1.43truein]{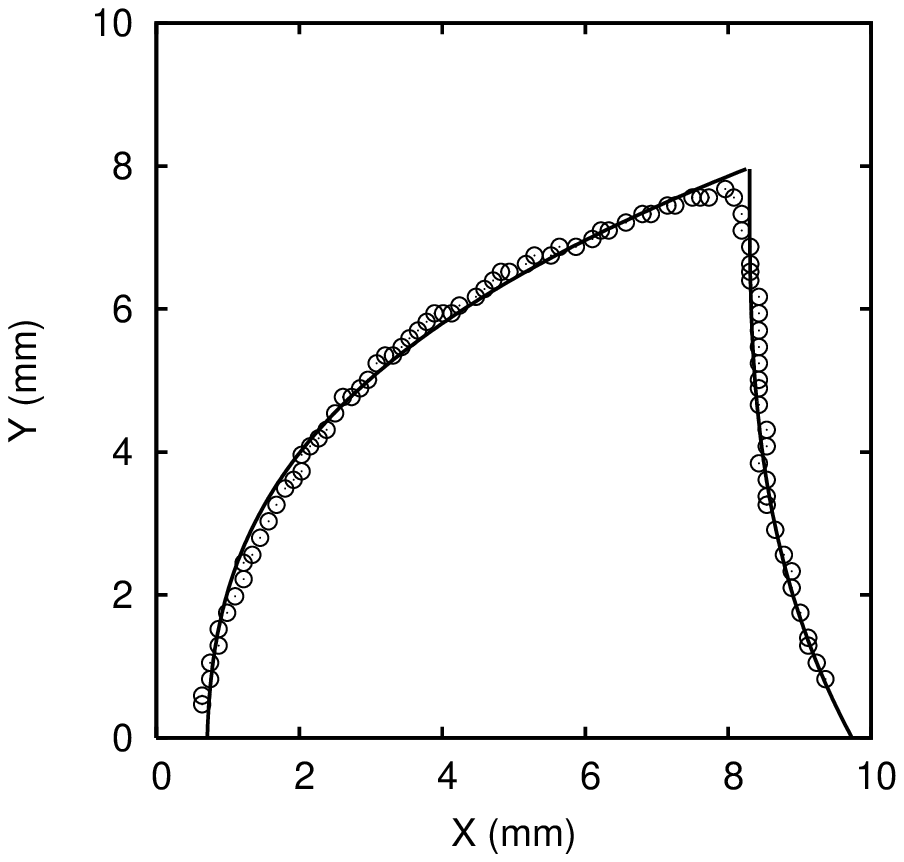}
\caption{\label{f-tear}
Detail of the tear shape, for a $d = 9.5$ mm acrylic cylinder, $U_0 =$ 77 mm/s, with cubic arc fit (see text).}
}
\end{figure}

The striking tearing shape 
(Fig.~\ref{f-tear}a) occurs because the crack propagates initially perpendicular to the cavity surface, but curves back as the cylinder advances and the tear opens up. A similar geometric effect is responsible for the shapes observed in the tearing of mylar by a moving cylinder \cite{roman03, ghatak03}. We find that this shape is best fit by a simple cubic arc $y = A(x - x_0)^3$ (Fig.~\ref{f-tear}b), where $x_0$ is the position directly behind the cylinder, rather than a cycloidal arc  \cite{ghatak03} or a parabolic arc  \cite{audoly05}.

The transition boundary from Flow to Cut follows a linear scaling $U_0 \sim d$, corresponding to a critical $De$ condition  $U_0 \lambda / d = De_\mathrm{c}$. Thus the onset of texturing occurs when the flow timescale $d/U_0$ becomes faster than the relaxation time $\lambda$; the material does not have time to relax. From the straight line in Fig.~\ref{phase} we obtain $De_c \simeq 1.1$.

The second transition occurs when the velocity of the cylinder is further increased, either from the Cutting state, or at larger diameters directly from the Flow state, as shown in Fig.~\ref{phase}. The higher velocity Tearing state is characterized by lateral cracks in the sidewalls of the wake  (see Figs.~\ref{fct_pics}b and \ref{f-tear}). The transition boundary for tearing seems to be hyperbolic ($U_0 d \sim$ constant), with a fitted constant of $31 \pm 4$ mm$^2$/s, as shown in Fig.~\ref{phase}.

%%%%%%%%%%%%%%%%%%%%%%%%%%%%%%%%%%%%
%%%%%%%%%%%%%%%%%%%%%%%%%%%%%%%%%%%%

The scaling of the tearing transition boundary can be derived in terms of viscoelastic crack propagation. The tearing involves what is essentially a {\it Mode I crack} \cite{freundBK}, propagating for a limited time 
in the cavity sidewall. The Griffiths condition for crack propagation is that the applied surface force per unit length 
(or energy per area) exceeds a critical value $\Gamma_\mathrm{crit}$ \cite{lawn93, tanaka00}. In our experiment however the cylinder speed $U_0$ is fixed, so we obtain an effective force using the approach of St.~Venant \cite{love,pastaPRL}:
assuming the resistance to the moving cylinder is that of an elastic solid, the force exerted on the cylinder is $F_d \sim E A U_0 /c$, where $E$ is Young's Modulus, $A = h d$ is the projected cross-sectional area in the direction of motion, and $c=\sqrt{E/\rho}$ is the longitudinal wavespeed \cite{love}. A crack in the cavity surface will propagate when the cylinder speed or size is such that $F_d$ exceeds $\Gamma_\mathrm{crit} h$, or
\begin{equation}
\frac{E \,d U_0}{c} > \Gamma_\mathrm{crit}\label{Etearcond}
\end{equation}
This leads to the hyperbolic boundary shown in Fig.~\ref{phase}.

Combining the measured phase diagram with Eq.~(\ref{Etearcond}) allows us to estimate the tear strength of our micellar material. 
We obtain the Young's modulus $E$ from our rheology: $E = 3G_0 \simeq 570$ Pa (assuming the incompressible Poisson ratio 1/2), thus $\Gamma_\mathrm{crit} \simeq 25 \pm 3$ mN/m.
This is significantly lower than typical tear strengths for 
elastomer gels (e.g. $\sim 10$ N/m \cite{gent96, tanaka00}), although it is close to some measured values for the surface free energy of soft materials (e.g. 20 - 25 mN/m for PDMS 
\cite{chaud96}). 
It is also not far from typical equilibrium surface tensions measured for 
wormlike micellar fluids: 30 - 36 mN/m \cite{cooper02,akers}.
Theoretically, the tear strength can be decomposed into surface energy and dissipative contributions \cite{gent96,persson05}, and the high tear strength of certain viscoelastic gels is attributed to a high dissipation  \cite{gent96,persson05}.
While it is not unusual for a soft solid to have a
surface free energy equal to its surface tension  \cite{chaud96} (both defined as the energy required to make new surface area), it is surprising that our measured tear strength $\Gamma_\mathrm{crit}$ is so close to the surface tension. This
indicates that dissipation processes are negligible for 
these slow cracks.

%%%%%%%%%%%%%%%%%%%%%%%%%%
\begin{figure}
\centering
\includegraphics[width=3.2truein]{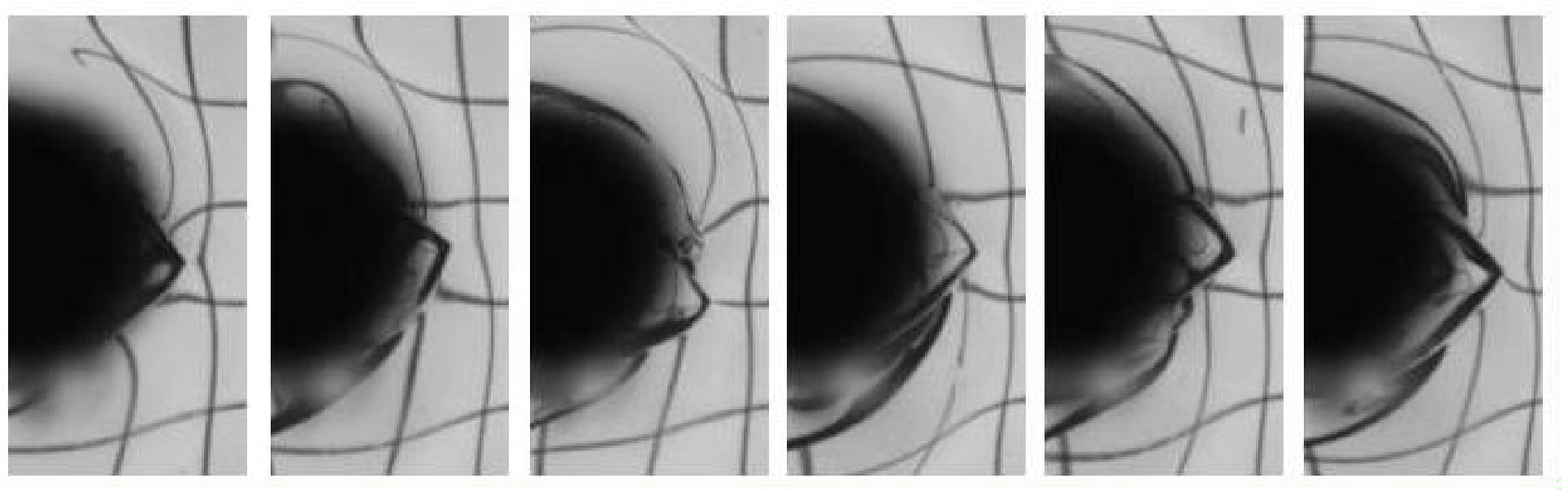}\\
\vspace{0.5cm}
\includegraphics[width=2.7truein]{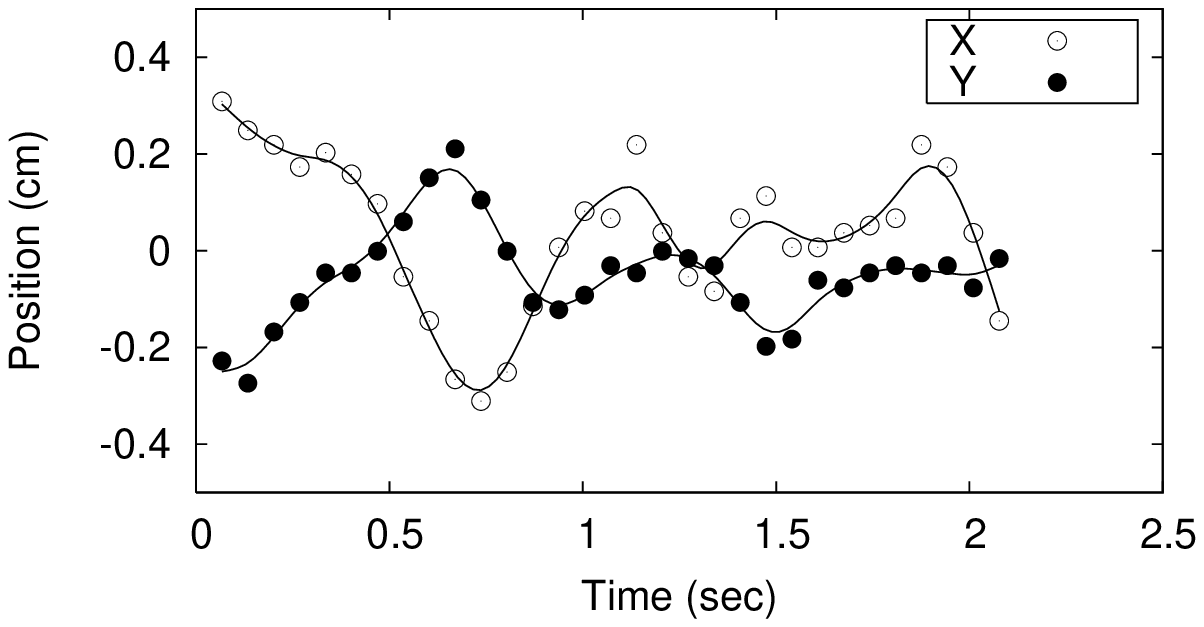}
\caption{Oscillating driven crack preceding the cylinder  ($d = 9.5$ mm, $U_0 =$ 26 mm/s). The plot shows the horizontal ($X$) and vertical ($Y$) positions of the crack tip.
The origin is arbitrarily offset to a point in front of the cylinder and in line with its motion. The solid lines are 
splines to guide the eye.
}
\label{crack}
\end{figure}
%%%%%%%%%%%%%%%%%%%%%%%%%%

At high speeds in the tearing regime, a crack is observed {\it in front of} the cylinder. This leading crack is unsteady, swaying side-to-side, sometimes losing out to another crack, as shown in Fig.~\ref{crack}. The oscillation of both the transverse ($Y$) and longitudinal ($X$) directions is similar to the dynamics of cracks driven in similar experiments using mylar sheets \cite{roman03, ghatak03}.
With an opening angle $\sim 100^{\circ}$, the crack tip we observe is not the needle shape seen in brittle materials \cite{lawn93}. Wedge-shaped crack tips are also seen in  rubber \cite{deegan}, and are characteristic of soft materials where a large deformation is needed to achieve the stress levels for 
propagation \cite{marderPRL}.
However these cracks are not driven by remote loading as in Mode I crack propagation \cite{freundBK,lawn93}, but by local contact with the moving cylinder. Thus the crack speed follows the cylinder speed $U_0$; for our material $c \simeq 72$ cm/s, and $U_0/c \simeq 0.05$.
A similar slow driven fracture was seen in peeling experiments on gels \cite{tanaka00}.

Wormlike micellar fluids exhibit a strong flow birefringence so that when viewed through cross polarizers, the stress field can be readily visualized  \cite{hu93,handzy04}; a feature known as the photoelastic effect \cite{dally91}.
In a birefringent image, monochromatic fringes represent isolines for the stress field, and the spacing is proportional to the stress intensity. For the experiments reported here we find a characteristic dipole pattern around the cylinder, with a pronounced asymmetry due to the stress relaxation in the wake \cite{IUTAMexico}.
The fringes are most closely spaced at the sides of the cylinder, where the shear deformation is maximum; this is also where the onset of the cutting instability occurs  \cite{IUTAMexico}. Surface interactions may also play a role in this transition, analogous to the sharkskin instability in polymer melt extrusion (see e.g.~\cite{dennANNREV}).

%%%%%%%%%%%%%%%%%%%%%%%%%%%%%%%%%%%%%%%%%%%%%%
\begin{figure}
\centering
\includegraphics[width=3.2truein]{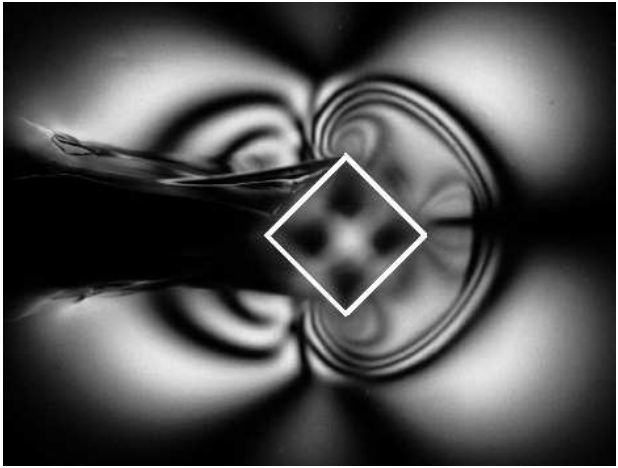}
\caption{Birefringent visualization of the micellar fluid layer around a diagonally oriented square `cutting tool' ($S =$ 12.5 mm, $U_0 = 2$ mm/s to the right, depth $H =  0.9$ cm). A white outline is superimposed on the outer edge of the square.}
\label{Fsquare}
\end{figure}
%%%%%%%%%%%%%%%%%%%%%%%%%%%%%%%%%%%%%%%%%%%%%%

Cutting 
is however not usually done with a cylinder. We expected that a rod with sharper corners - something like a knife - would focus the stress at the sharp front edge and cut more directly. Thus we were surprised to find, in a series of similar experiments using a square bar, that the high stress started at the {\it side corners}, and formed an apparent detached stress boundary layer which travelled ahead of the square, evident in Fig.~\ref{Fsquare} for side length $S = 12.5$ mm and $U_0 = 2$ mm/s  ($De \simeq 0.18$).
The square did cut more efficiently, in the sense that all instabilities (Cut/Tear) occurred at higher speeds than for the cylinder, much as opening an envelope with a finger leaves ragged edges (tearing) while a letter opener at the same speed leaves a cleaner edge \cite{ghatak03}.

In conclusion, we have observed transitions from fluid flow past a cylinder to solid-like tearing and fracture in a single experiment with concentrated micellar fluid.
This soft material, comprised of surfactant aggregates weakly held together by the hydrophobic effect \cite{israelBK}, sits at the material limit of weak resistance to tear; the measured tear strength is approximately equal to the surface tension.
At the highest speeds we observe a driven crack in front of the moving cylinder.
Unlike standard crack propagation, in which the experiments use remote loading, here the viscoelastic crack is driven at the speed of the cylinder $U_0$. 
It remains to be seen whether any drastic changes in the crack dynamics occur as $U_0$ increases towards a natural 
speed 
threshold of the material, perhaps the elastic wave speed. 
What this study has shown is that - from slow crack propagation to strong birefringence to tearing and ``self-healing'' - concentrated micellar materials are an exemplary soft matter system, demonstrating the variety of complex dynamics possible in a simple combination of surfactant, salt, and water.

We would like to thank R. H. ``Bob'' Geist, B. Bitel, and D. Hill for experimental assistance, and E. Sharon, F. Costanzo, G. H. McKinley, A. Mazzucato, and J. Walton for helpful discussions. 
AB acknowledges support from the National Science Foundation (CAREER Award DMR-0094167).

%%%%%%%%%%%%%%%%%%%%%%%%%%%%%%%%%%%%%%%%%%%%%%
%%%%%%%%%%%%%%%%%%%%%%%%%%%%%%%%%%%%%%%%%%%%%%
%%%%%%%%%%%%%%%%%%%%%%%%%%%%%%%%%%%%%%%%%%%%%%

%%%%%%%%%%%%%%%%%%%%%%%%%%%%%%%%%%%%%%%%%%%%%%%%%%%
%%%%%%%%%%%%%%%%%%%%%%%%%%%%%%%%%%%%%%%%%%%%%%%%%%%

\end{document}